\documentclass[aps,apl,twocolumn]{revtex4}

\usepackage{amsmath,cases}
\usepackage{amsmath}
\usepackage{amssymb}
\usepackage[dvipdfmx]{graphicx}  
\usepackage[dvipdfmx]{color}
\usepackage{verbatim}   
\usepackage{color}

\newcommand{\bra}[1]{\langle{#1}|}
\newcommand{\ket}[1]{|{#1}\rangle}

\begin{document}

\title{Quantum speed limit for robust state characterization and engineering}
\author{Kohei Kobayashi}
\author{Naoki Yamamoto}
\affiliation{Keio University, Department of Applied Physics and Physico-Informatics, 
Hiyoshi 3-14-1, Kohoku, Yokohama, 223-8522, Japan}
\date{\today}

\begin{abstract}

In this paper, we propose a concept to use a quantum speed limit (QSL) as a measure of 
robustness of states, defining that a state with bigger QSL is more robust. 
In this perspective, it is important to have an explicitly-computable QSL, because then we 
can formulate an engineering problem of Hamiltonian that makes a target state robust 
against decoherence. 
Hence we derive a new explicitly-computable QSL that is applicable to general Markovian 
open quantum systems. 
This QSL is tighter than another explicitly-computable QSL, in an important setup 
such that decoherence is small. 
Also the Hamiltonian engineering problem with this QSL is a quadratic convex optimization 
problem, and thus it is efficiently solvable. 
The idea of robust state characterization and the Hamiltonian engineering, in terms of QSL, 
is demonstrated with several examples. 

\end{abstract}

\maketitle


\section{Introduction}

{\it Quantum speed limit} (QSL) is a lower bound on the evolution time of a quantum system 
from an initial state to a final state. 
It has numerous applications in quantum computation \cite{computation1, computation2}, 
metrology \cite{metrology1, metrology2}, optimal control \cite{opt1, opt2, opt3}, and so on. 
The first study of QSL was focused on closed systems; 
Mandelstam and Tamm derived a QSL between orthogonal states, which is given by the 
variance of Hamiltonian \cite{MT}, and Margolus and Levitin derived another QSL represented 
by the mean energy \cite{ML}. 
Moreover, the extensions to mixed states \cite{Uhlmann} and time-dependent driven systems 
\cite{Pfeifer1, Pfeifer2, Deffner2} were presented later. 
In recent years, several type of QSLs for open quantum systems \cite{Taddei, Campo, Deffner, 
Meng, Sun, Zhang, Campaioli, speed1, speed2, speed3, speed4, speed5} have been 
extensively investigated.

In this paper, we exploit a new application of QSL; 
that is, we use QSL to characterize robust quantum states of a given open quantum system. 
Typically, QSL is used to characterize the potential for speeding up the time evolution toward 
a target state \cite{Meng, speed1, speed2, speed3, speed4, speed5}. 
More precisely, let us consider the problem to transfer an initial state $\rho_0$ to a target 
final state $\rho_f$; 
if the QSL from $\rho_0$ to $\rho_f$ of a system $\Sigma$ is smaller than that of 
another system $\Sigma'$, then $\Sigma$ should be chosen to do this task. 
In contrast, in this paper we consider an undesired state evolution driven by decoherence. 
That is, we consider a QSL from $\rho_0$ to any state $\rho_T$ such that the distance 
between $\rho_0$ and $\rho_T$ is bigger than a certain fixed value. 
If this QSL is large, this means that the decoherence needs a lot of time to drive the state 
initialized at $\rho_0$ toward $\rho_T$; 
in other words, $\rho_0$ is not largely affected by the decoherence. 
In this view, therefore, $\rho_0$ with a large QSL is robust against the decoherence.

Based on the above-mentioned use of QSL, we consider the following optimization problem; 
the goal is to engineer the system Hamiltonian that maximizes the QSL for a given $\rho_0$ 
and the decoherence. 
Note that, to make this optimization problem tractable, it is important that the QSL has an 
explicit expression in terms of the parameters, rather than an implicit one that needs, for 
instance, solving a differential equation. 
Actually in this paper we derive a new easy-to-compute QSL applicable to a general Markovian 
open quantum system and prove that it is tighter than another explicit QSL given in 
Ref.~\cite{Campo}, in the setup where the decoherence strength and the distance are 
both small. 
Moreover, it is shown that the Hamiltonian engineering problem based on this new QSL is 
a quadratic convex optimization problem, which is efficiently solvable.


\section{New explicit quantum speed limit}

\subsection{Setup and derivation}

In this paper, we consider the general open quantum system obeying the Markovian master 
equation 
\begin{eqnarray}
    \frac{d\rho_{t}}{dt}=-i[H, \rho_{t}]+\mathcal{D}[M]\rho_t,
\end{eqnarray}
where $H$ is the time-independent Hamiltonian and $\mathcal{D}[M]$ is the Lindblad 
superoperator defined by 
$\mathcal{D}[M]\rho=M\rho M^{\dagger}-M^{\dagger}M\rho/2-\rho M^{\dagger}M/2$. 
Throughout the paper, we assume that the initial state is pure; 
$\rho_{0}=\ket{\psi_{0}}\bra{\psi_{0}}$. 
Next, following \cite{Campo, Meng, Zhang}, we define the relative purity between $\rho_{0}$ 
and $\rho_{t}$ as 
\begin{equation}
      \Theta_{t} = {\rm arccos}\left\{{\rm Tr}(\rho_{0}\rho_{t})\right\}.
\end{equation}
Clearly, $0\leq \Theta_t \leq \pi/2$. 
This takes the maximum when $\rho_{t}$ is orthogonal to $\rho_{0}$, and the minimum is 
achieved only when $\rho_t = \rho_0$. 
Hence, the relative purity can be interpreted as a distance between $\rho_{0}$ and $\rho_{t}$. 
Here we derive a new lower bound of the time $T$, needed for the relative purity to evolve 
from $\Theta_{0}=0$ to a given $\Theta_{T}\in(0, \pi/2]$.

First, we find that the dynamics of $\Theta_{t}$ is given by
\begin{align}
\label{theorem proof eq 1}
     \frac{d\Theta_{t}}{dt}
         &=\frac{-1}{\sqrt{1-{\rm Tr}(\rho_{0}\rho_{t})^{2}}} 
                     \cdot {\rm Tr}\Big( \rho_{0} \frac{d\rho_{t}}{dt} \Big) 
\notag \\
         &=\frac{1}{{\rm sin}\Theta_{t}} 
              {\rm Tr}\Big\{ \Big( i[\rho_{0}, H]- \mathcal{D}^{\dagger}[M]\rho_0 \Big)\rho_t \Big\},
\end{align}
where $\mathcal{D}^{\dagger}[M]\rho
=M^{\dagger}\rho M-M^{\dagger}M\rho/2-\rho M^{\dagger}M/2$. 
To have an upper bound of the rightmost side of Eq.~\eqref{theorem proof eq 1}, we use two 
inequalities. 
One is the Cauchy-Schwarz inequality for matrices $X$ and $Y$: 
\begin{equation}
\label{Cauchy-Schwarz} 
     \left|{\rm Tr}(X^{\dagger}Y)\right| \leq \|X\|_{\rm F}\|Y\|_{\rm F},
\end{equation}
where $\|X\|_{{\rm F}}=\sqrt{{\rm Tr}(X^{\dagger}X)}$ is the Frobenius norm. 
The other one is as follows; 
\begin{align*}
     \|\rho_{t}-\rho_{0}\|_{\rm F} 
         &=\sqrt{{\rm Tr}\big[\left(\rho_{t} -\rho_{0}\right)^{2}\big]} 
           =\sqrt{{\rm Tr}\left( \rho_{t}^2-2\rho_{t} \rho_{0}+\rho_{0}^{2}\right)} 
\\
         & \leq \sqrt{2-2{\rm Tr}(\rho_{t}\rho_{0})} =\sqrt{2-2 \cos \Theta_{t}} ,
\end{align*}
where ${\rm Tr}(\rho_{t}^{2})\leq 1$ and ${\rm Tr}(\rho_{0}^{2})=1$ are used. 
Using these inequalities, the rightmost side of Eq.~\eqref{theorem proof eq 1} is upper 
bounded by 
\begin{align}
\label{theorem proof eq 2}
     & {\rm Tr}\{ \left(i[\rho_{0}, H] -  \mathcal{D}^{\dagger}[M]\rho_0 \right) \rho_{t} \}  
\notag \\
     & ={\rm Tr}\{ \left(i[\rho_{0}, H]- \mathcal{D}^{\dagger}[M]\rho_0 \right)(\rho_{t}-\rho_{0})\}
           - {\rm Tr}(\rho_{0} \mathcal{D}^{\dagger}[M]\rho_0 ) 
\notag\\
     &={\rm Tr}\{ \left(i[H, \rho_{0}]+ \mathcal{D}^{\dagger}[M]\rho_0 \right)(\rho_{0}-\rho_{t})\} 
           + {\rm Tr}(M^{\dagger}M\rho_{0}) 
\notag\\
      &\ \ \ - {\rm Tr}(M^{\dagger}\rho_{0}M\rho_{0}) 
\notag\\
      & \leq \|i[H, \rho_{0}]  + \mathcal{D}^{\dagger}[M]\rho_0 \|_{\rm F} \cdot 
              \|\rho_{t}-\rho_{0}\|_{\rm F} + \|M\ket{\psi_{0}}\|^{2} 
\notag \\
      &\ \ \ -|\bra{\psi_{0}}M\ket{\psi_{0}}|^{2}  
\notag \\
      &\leq \sqrt{2}\| i[H, \rho_{0}] + \mathcal{D}^{\dagger}[M]\rho_0 \|_{\rm F}
            \sqrt{1-{\rm cos}\Theta_{t}} + \|M\ket{\psi_{0}}\|^{2} 
\notag\\
      & \ \ \ -|\bra{\psi_{0}}M\ket{\psi_{0}}|^{2},  
\end{align}
where $\|\ket{\psi_{0}}\|^{2}=\langle\psi_{0}|\psi_{0}\rangle$ is the Euclidean norm. 
From Eqs.~\eqref{theorem proof eq 1} and \eqref{theorem proof eq 2}, we have 
\begin{eqnarray}
\label{theorem proof eq 3}
      \frac{d\Theta_{t}}{dt} 
          \leq \frac{1}{{\rm sin}\Theta_{t}}
               \left( \mathcal{A}\sqrt{1-{\rm cos}\Theta_{t}}+\mathcal{E} \right),
\end{eqnarray}
where 
\begin{eqnarray*}
     \mathcal{A}&=&\sqrt{2}\|i[H, \rho_{0}]+ \mathcal{D}^{\dagger}[M]\rho_0 \|_{\rm F},  
\\
     \mathcal{E}&=&\|M\ket{\psi_{0}}\|^{2}-|\bra{\psi_{0}}M\ket{\psi_{0}}|^{2}.
\end{eqnarray*}
Then by integrating the inequality \eqref{theorem proof eq 3}, from $0$ to $T$, we end up with 
\begin{align}
\label{T star}
    T \geq T_{\ast}(\rho_0) := \frac{2  \lambda }{\mathcal{A}} 
            +\frac{2\mathcal{E}}{\mathcal{A}^{2}}
                   \ln \left( \frac{\mathcal{E}}{\mathcal{E}+\mathcal{A}\lambda  }\right),  
\end{align}
where $\lambda=\sqrt{1- \cos\Theta_{T}}$. 
We often write simply $T_*$ rather than $T_{\ast}(\rho_0)$. 
This $T_{\ast}$ is our QSL, giving a lower bound on the evolution time $T$ for the state 
$\rho_t$ to evolve from $\rho_0$ to any state $\rho_T$ satisfying 
$\Theta_T = {\rm arccos}\left\{{\rm Tr}(\rho_0\rho_T)\right\}$ for a given value of $\Theta_T$.

Here we list the points of $T_{\ast}$. 
\\
\ (i) $T_{\ast}$ is explicitly represented in terms of $(\rho_0, H, M, \Theta_T)$, and thus it is 
readily computable once those parameters are specified. 
There is no need to solve any equation. 
\\
\ (ii) $T_{\ast}$ is monotonically decreasing with respect to the magnitude of $M$; see 
Appendix A for the proof. 
This implies that, as the decoherence becomes bigger, the dynamical change of state can 
become faster. 
\\ 
\ (iii) $T_{\ast}$ is monotonically decreasing with respect to $\mathcal{A}$ for a fixed 
$\mathcal{E}$; see Appendix B for the proof. 
In general, a closed system with bigger Hamiltonian $H$ evolves faster; but in the case of 
open quantum systems, this effect may be changed by the decoherence effect $M$. 
Intuitively, $\mathcal{A}$ corresponds to the amplitude of such an effective Hamiltonian. 
Later in Section IV, we will see that the monotonically decreasing property of $T_{\ast}$ 
with respect to $\mathcal{A}$ is used to formulate the Hamiltonian engineering problem 
for robust state generation. 
\\
\ (iv) It is straightforward to extend the result to the case where the system is subjected to 
multiple decoherence channels and Hamiltonians. 
In this case $T_{\ast}$ is given by Eq.~\eqref{T star} with 
\begin{align*}
      \mathcal{A} &= \sqrt{2} \Big\| \sum_{j}i[H_{j}, \rho_{0}]
               +\sum_{j} \mathcal{D}^{\dagger}[M_j]\rho_0   \Big\|_{\rm F}, 
\\
      \mathcal{E}&=\sum_{j}\left(\|M_{j}\ket{\psi_{0}}\|^{2} 
               - |\bra{\psi_{0}}M_{j}\ket{\psi_{0}}|^{2}\right).  
\end{align*}
%


\subsection{Quantum speed limit as a measure of robustness}
\label{sec:robust characterization}

\begin{figure}[htbp]
\includegraphics[width=6cm]{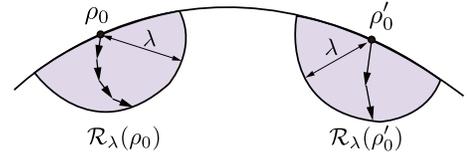}
\caption{Robustness of the quantum state. 
(Left) When $T_{\ast}(\rho_0)$ is large, the quantum state $\rho_t$ must take a long time to 
exit from the region ${\mathcal R}_\lambda(\rho_0)$ with fixed radius $\lambda$. 
This implies that $\rho_{0}$ is robust. 
(Right) When $T_{\ast}(\rho_0')$ is small for the same $\lambda$, then $\rho_t$ may quickly 
exit from ${\mathcal R}_\lambda(\rho_0')$, meaning that $\rho_{0}'$ is fragile compared 
to $\rho_0$. 
}
\label{robust figure}
\end{figure}

Let us consider the situation where an initial state $\rho_0=\ket{\psi_0}\bra{\psi_0}$ and 
a value of $\lambda=\sqrt{1-\cos\Theta_T}$ are given. 
This means that we are given a region ${\mathcal R}_\lambda(\rho_0)$, which is the set 
of all states whose distance from $\rho_0$ is less than $\lambda$. 
That is, $\lambda$ can be interpreted as the radius of a circle region 
${\mathcal R}_\lambda(\rho_0)$; see Fig.~\ref{robust figure}. 
Then the transition time $T$ of $\Theta_t$ for evolving from $\Theta_0=0$ to $\Theta_T$ 
has the meaning of the escape time that the state first exits from ${\mathcal R}_\lambda(\rho_0)$. 
Therefore, if $T$ is large for a given $\lambda$, this means that the state $\rho_t$ starting 
from the initial state $\rho_0$ takes a long time to exit from ${\mathcal R}_\lambda(\rho_0)$. 
In this case, we can say that $\rho_0$ is robust against the decoherence $M$. 
In contrast, if we take another initial state $\rho_0'$ and find that the transition time $T'$ is 
smaller than $T$ for the same value of $\lambda$, this means that the state quickly escapes 
from ${\mathcal R}_\lambda(\rho_0')$; 
that is, as illustrated in Fig.~\ref{robust figure}, this is the case where the state $\rho_0'$ 
is largely affected by the decoherence and can be easily changed. Thus, $\rho_0'$ is fragile.

Hence it is clear that the QSL $T_*(\rho_0)$ can be used to characterize a state $\rho_0$ 
that is robust against a given decoherence $M$. 
That is, for a given $\lambda$, the state initialized to $\rho_0$ with a large value of $T_*(\rho_0)$ 
is guaranteed to take a long time $T$ to escape from ${\mathcal R}_\lambda(\rho_0)$, hence it is 
robust against $M$. 
Also, in this paper we define that, if $T_*(\rho_0)>T_*(\rho_0')$, then $\rho_0$ is more robust 
than $\rho_0'$, although this does not always lead to $T(\rho_0)>T(\rho_0')$. 
Moreover, for a given $\rho_0$ and (a relatively small value of) $\lambda$, it makes sense 
to appropriately design the system operators that maximize $T_{\ast}(\rho_0)$, to protect 
$\rho_0$ against the decoherence; 
in Section IV we discuss this problem, especially in the case where $H$ is the design 
object.


\subsection{Comparison to the QSL derived in Ref.~\cite{Campo}}
\label{sec:comparison}

Applying the Cauchy-Schwarz inequality \eqref{Cauchy-Schwarz} to the right-hand side of 
Eq.~\eqref{theorem proof eq 1}, we have 
\[
     \frac{d\Theta_{t}}{dt} \leq \frac{1}{\sin\Theta_t} 
              \| i[\rho_{0}, H]- \mathcal{D}^{\dagger}[M]\rho_0 \|_{\rm F} 
                   \| \rho_t \|_{\rm F} 
             \leq \frac{\mathcal{A}}{\sqrt{2}\sin\Theta_t}. 
\]
Then by integrating both sides of this inequality from $0$ to $T$, we find that the transition 
time $T$ for $\Theta_t$ evolving from $\Theta_0=0$ to a given $\Theta_T$ is lower bounded 
by \cite{Campo}: 
\begin{align} 
        T \geq T_{\rm DC}
          :=\frac{\sqrt{2}(1-\cos \Theta_{T}) }{ \mathcal{A} } 
          =\frac{\sqrt{2}\lambda^{2}}{\mathcal{A}}.
\end{align}
Similar to $T_{\ast}$, $T_{\rm DC}$ is also explicitly represented in terms of 
$(\rho_0, H, M, \Theta_T)$, which is indeed the key point for engineering a system having 
a robust state $\rho_0$ in the sense of QSL as described in 
Section~\ref{sec:robust characterization}. 
Note that, to our best knowledge, no explicit form of QSL for open quantum systems 
has been developed, except for $T_{\ast}$ and $T_{\rm DC}$.

Therefore, it is important to compare $T_*$ and $T_{\rm DC}$. 
We study the following quantity: 
\begin{align}
\label{VS DC bound}
    \frac{T_{\ast}}{T_{\rm DC}} 
        &=\left\{ \frac{2  \lambda }{\mathcal{A}} 
           + \frac{2\mathcal{E}}{\mathcal{A}^{2}}
               \ln{\left(\frac{\mathcal{E}}{\mathcal{E}+\mathcal{A}\lambda  }\right)}  \right\}
               \frac{\mathcal{A}}{\sqrt{2}\lambda^{2}} 
\nonumber \\
        &= \frac{ \sqrt{2}}{\lambda} 
              + \frac{\sqrt{2}k}{\lambda^{2}}\ln\left( \frac{k}{k+\lambda} \right),
\end{align}
where $k=\mathcal{E}/\mathcal{A}$. 
Note that again from Eq.~\eqref{Cauchy-Schwarz} we have 
\begin{align*}
     \mathcal{A} &= \sqrt{2}\| i[H, \rho_{0}] + \mathcal{D}^{\dagger}[M]\rho_0 \|_{\rm F} 
                                 \cdot \| \rho_{0} \|_{\rm F}
\\
     & \geq \sqrt{2} | {\rm Tr}\{ ( i[H, \rho_{0}] + \mathcal{D}^{\dagger}[M]\rho_0 ) \rho_0 \} |
     = \sqrt{2} \mathcal{E}, 
\end{align*}
hence $0\leq k \leq 1/\sqrt{2}$. 
First, $T_{\ast}/T_{\rm DC}$ is a monotonically decreasing function with respect to $k$, because 
\begin{align*}
      \frac{\partial}{\partial k}\left(\frac{ T_{\ast}}{T_{\rm DC}} \right)
         &=\frac{\sqrt{2}}{\lambda} \left\{ - \frac{1}{\lambda}\ln\left(1+\frac{\lambda}{k} \right)
                   + \frac{1}{k+\lambda}   \right\}  \\
         &\leq \frac{\sqrt{2}}{\lambda}\left(-\frac{1}{\lambda} \frac{\lambda}{k+\lambda}  
                  + \frac{1}{k+\lambda}    \right) =0,
\end{align*}
where we used $\ln(1+x) \geq x/(1+x)$ for $x\geq0$. 
Now, when $k=0$ or equivalently when the system is closed (i.e., $\mathcal{E}=0$), then 
$T_{\ast}/T_{\rm DC} \geq \sqrt{2}/\lambda > 1$. 
Together with the above monotonically decreasing property of $T_{\ast}/T_{\rm DC}$ with 
respect to $k$, hence, $T_*$ is tighter than $T_{\rm DC}$ if the decoherence is small.

Next, $T_{\ast}/T_{\rm DC}$ decreases with respect to $\lambda$, because
\begin{align*}
     \frac{\partial}{\partial \lambda}\left(\frac{ T_{\ast}}{T_{\rm DC}} \right)
         &= -\frac{\sqrt{2}}{\lambda^2}+\frac{2\sqrt{2}k}{\lambda^3}
                  \ln \left(1+\frac{\lambda}{k}\right)
                  -\frac{\sqrt{2}k}{\lambda^2}\frac{1}{k+\lambda} \\
         & \leq \frac{\sqrt{2}}{\lambda^2}
                 \left\{ -1+\frac{2k+\lambda}{k+\lambda}
                     -\frac{k}{k+\lambda} \right\} =0,
\end{align*}
where we used $\ln(1+x) \leq x(2+x)/2(1+x)$ for $x\geq0$. 
This means that $T_*$ will work as a tighter bound than $T_{\rm DC}$, in the region 
${\mathcal R}_\lambda(\rho_0)$ with small radius $\lambda$.

\begin{figure}[htbp]
\includegraphics[width=6cm]{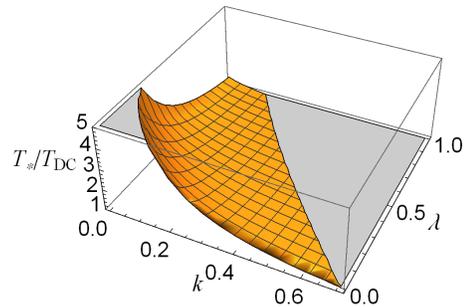}
\caption{
The 3-dimensional plot of the ratio of bounds, $T_{\ast}/T_{\rm DC}$, as a function of 
$k\in[0, 1/\sqrt{2}]$ and $\lambda\in[0,1]$. 
}
\label{comparison figure}
\end{figure}

The above observations can be quantitatively seen in Fig.~\ref{comparison figure}, which 
plots Eq.~\eqref{VS DC bound} 
as a function of $k\in[0, 1/\sqrt{2}]$ and $\lambda\in[0,1]$. 
The yellow-colored region shows the set of parameters $(\lambda, k)$ such that 
$T_{\ast} > T_{\rm DC}$. 
Notably, when the decoherence is weak (i.e., $k$ is small) and ${\mathcal R}_\lambda(\rho_0)$ 
is small (i.e., $\lambda$ is small), then $T_*$ functions as a much bigger lower bound for the 
escape time $T$, than $T_{\rm DC}$.


\section{Examples}
\label{sec:examples}

\subsection{Two-level atom}

The first example is a two-level atom consisting of the excited state $\ket{0} = [1, 0]^{\top}$ 
and the ground state $\ket{1}=[0, 1]^{\top}$. 
Let the initial state $\ket{\psi_{0}}$ be 
\begin{equation*}
     \ket{\psi_{0}} 
          = [{\rm cos}\theta, e^{i\varphi}{\rm sin}\theta]^{\top}, \ 
              (0\leq \theta < \pi/2,\  0\leq \varphi < 2\pi).
\end{equation*}
We consider the following system operators: 
\begin{equation}
\label{qubit example 1}
     H=\omega\sigma_{z}, \ M=\sqrt{\gamma}\sigma_{x},
\end{equation}
where $\sigma_{x}=\ket{0}\bra{1}+\ket{1}\bra{0}$, $\sigma_{y}=i(\ket{1}\bra{0}-\ket{0}\bra{1})$, 
and $\sigma_{z}=\ket{0}\bra{0}-\ket{1}\bra{1}$ are the Pauli matrices. 
$H$ rotates the state vector along the $z$-axis with frequency $\omega>0$. 
$M$ represents the dephasing noise with decay rate $\gamma>0$. 
In this setting the QSL is given by Eq.~\eqref{T star} with 
\begin{align*}  
      \mathcal{A}^{2}/4 
         &=\gamma^{2}({\rm cos}^{2}2\theta+{\rm sin}^{2}2\theta{\rm sin}^{2}\varphi) 
             + \omega^{2}{\rm sin}^{2}2\theta \\
         &\ \ +\omega\gamma{\rm sin}^{2}2\theta{\rm sin}2\varphi, \\
      \mathcal{E} &=\gamma-\gamma{\rm sin}^{2}2\theta{\rm cos}^{2}\varphi.
\end{align*}

\begin{figure}[tb]
\includegraphics[width=8.8cm]{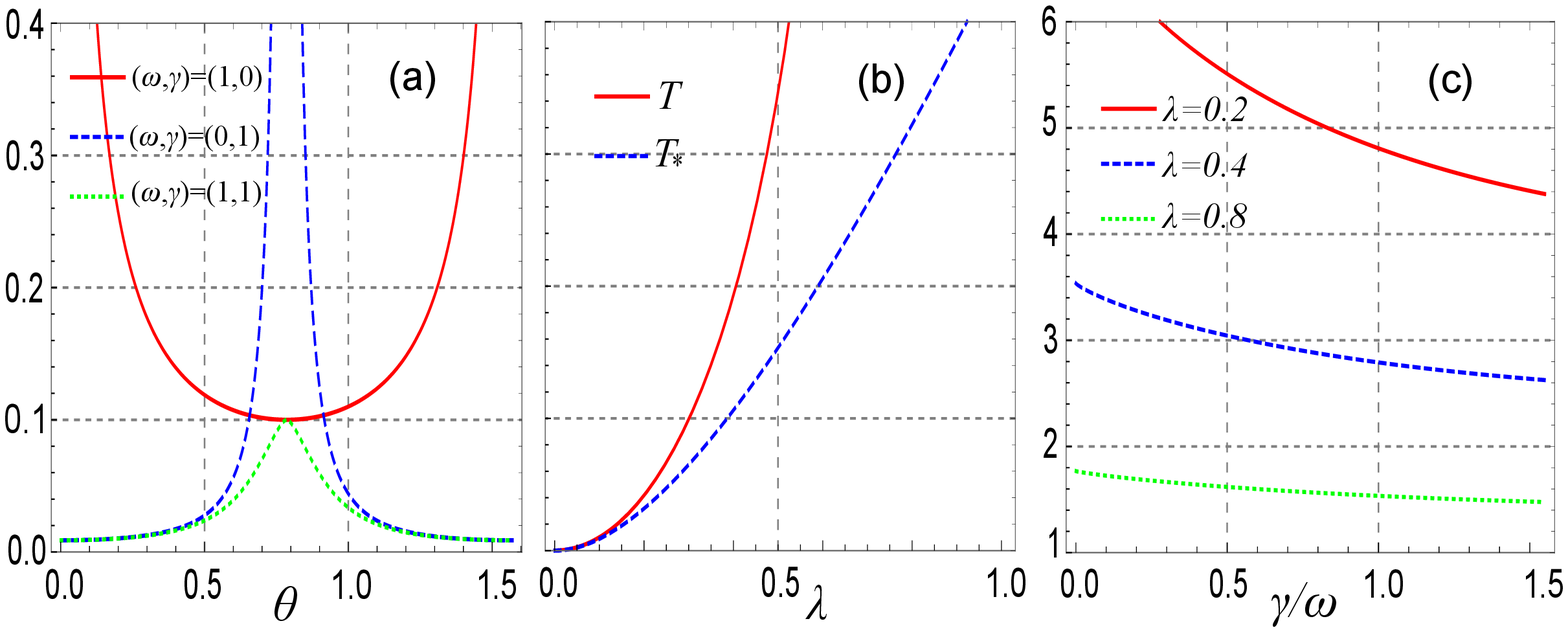}
\caption{
(a) The lower bound $T_*$ as a function of $\theta$, for several values of $(\omega, \gamma)$. 
For all cases, $\lambda=0.1$. 
(b) Comparison of the exact escape time $T$ and its lower bound $T_{\ast}$, as a function 
of $\lambda$, where $\gamma=1$ is fixed. 
(c) The ratio $T_{\ast}/T_{\rm DC}$, as a function of $\gamma / \omega$, for several values 
of $\lambda$. 
 }
 \label{qubit figure}
\end{figure}

Figure~\ref{qubit figure}(a) shows $T_{\ast}$ for the initial state with $\varphi=0$, as 
a function of $\theta$, for a fixed value $\lambda=0.1$. 
That is, this figure shows the lower bound of the escape time that the state initialized at 
$\ket{\psi_{0}} = [\cos\theta, \sin\theta]^{\top}$ first exits from the region 
${\mathcal R}_\lambda(\rho_0)$. 
If $\gamma=0$, $T_{\ast}=\lambda/\omega|{\rm sin}2\theta|$, which is plotted with the 
red solid line; 
in this case the state simply rotates along $z$ axis, and hence, if $\ket{\psi_{0}}$ is nearly 
$\ket{0}$ or $\ket{1}$, the state remains inside ${\mathcal R}_\lambda(\rho_0)$ for all time, 
resulting $T_*\rightarrow \infty$. 
Also $T_*$ takes the minimum at $\theta=\pi/4$, simply because the state on the equator 
of the Bloch sphere changes the most; 
hence $\ket{+}:=(\ket{0}+\ket{1})/\sqrt{2}$ is the most fragile state in our definition.

When $\gamma > 0$, the dependence of $T_{\ast}$ on $\theta$ remarkably changes, as 
shown with the blue dashed and green dotted lines in Fig.~\ref{qubit figure}(a). 
Again, $\lambda=0.1$ is chosen. 
When $(\omega,\gamma)=(0,1)$, $T_{\ast}\to\infty$ at $\theta=\pi/4$; that is, $\ket{+}$ 
is the most robust. 
Actually in this case, $\ket{+}$ is a steady state of the master equation 
$d\rho_t/dt = {\mathcal D}[M]\rho_t$, meaning that $\ket{+}$ does not change under the 
influence of this decoherence and the state around $\ket{+}$ remains in 
${\mathcal R}_\lambda(\rho_0)$ for all time. 
On the other hand, when $(\omega,\gamma)=(1,1)$, $T_{\ast}$ takes a finite time for 
all $\theta$, which implies that the state may escape from ${\mathcal R}_\lambda(\rho_0)$ 
at a certain time, for any $\rho_0$.

Recall that $T_*$ is a lower bound of the exact escape time $T$. 
Hence, it is worth comparing these quantities to see the tightness of $T_*$. 
For this purpose, here we set $\omega=0$ and choose the initial state $\ket{\psi_{0}} = \ket{0}$. 
In this case the master equation $d\rho_t/dt = {\mathcal D}[M]\rho_t$ yields a simple solution 
$\cos\Theta_{T}=(1+e^{-2\gamma T})/2$. 
As a result, we obtain $T$ and $T_{\ast}$ as follows: 
\begin{align*}
    T=-\frac{1}{2\gamma}\ln(1-2\lambda^2), ~~
    T_{\ast}=\frac{\lambda}{\gamma} - \frac{1}{2\gamma}\ln(2\lambda+1).
\end{align*}
Figure~\ref{qubit figure}(b) shows the plots of $T$ and $T_{\ast}$ with $\gamma=1$, 
as a function of $\lambda$, where the range of the vertical axis is the same as that of 
Fig.~\ref{qubit figure}(a). 
This shows that both $T$ and $T_{\ast}$ are close to zero when $\lambda$ is small, which is 
reasonable because the state will take a short time to escape from a small region 
${\mathcal R}_\lambda(\rho_0)$. 
However, the gap between $T$ and $T_{\ast}$ quickly diverges, as $\lambda$ becomes large. 
This fact suggests us to use $T_*$, only when $\lambda$ is small.

Lastly, let us see the ratio $T_{\ast}/T_{\rm DC}$ discussed in Sec.~\ref{sec:comparison}, 
particularly in the following setup; 
\begin{align*}
    H=\omega\sigma_{z}, \ M=\sqrt{\gamma}\sigma_{-}=\sqrt{\gamma}\ket{1}\bra{0},
\end{align*}
where $M$ represents the energy decay of the two-level atom with decay rate $\gamma>0$. 
The initial state is set to the superposition $\ket{\psi_0}=\ket{+}$. 
In fact, this example demonstrates the difference of the two lower bounds more drastically than 
the setting \eqref{qubit example 1}. 
We now have $\mathcal{A}=\sqrt{48\omega^2+11\gamma^2}/4$ and $\mathcal{E}=\gamma/16$. 
Then $T_{\ast}/T_{\rm DC}$ given in Eq.~\eqref{VS DC bound} depends on only 
$\gamma / \omega$ and $\lambda=\sqrt{1-\cos\Theta_T}$. 
Figure~\ref{qubit figure}(c) shows the plot of $T_{\ast}/T_{\rm DC}$, as a function of 
$\gamma / \omega$, for several values of $\lambda$. 
As expected from the discussion in Sec.~\ref{sec:comparison}, $T_{\ast} /T_{\rm DC}$ 
increases as $\gamma / \omega$ becomes small for all $\lambda$, and also it becomes 
bigger for smaller $\lambda$. 
That is, $T_{\ast}$ is a tighter bound than $T_{\rm DC}$, if the decoherence is relatively 
small and the region ${\mathcal R}_\lambda(\rho_0)$ is small.

\subsection{Bell states}

Next, we study the Bell states defined by 
\begin{equation*}
     \ket{\Phi^{\pm}}=\frac{1}{\sqrt{2}}(\ket{0}\ket{0}\pm\ket{1}\ket{1}), ~~ 
     \ket{\Psi^{\pm}}=\frac{1}{\sqrt{2}}(\ket{0}\ket{1}\pm\ket{1}\ket{0}), 
\end{equation*}
which are maximally entangled states. 
Which state is the most robust under a given decoherence? 
As seen in the previous example, comparing $T_{\ast}$ of these states provides an answer 
to this natural question.
Here we take the collective noise modeled by $M=\sqrt{\gamma}(\sigma_{-}\otimes I+I\otimes\sigma_{-})$. 
Further,  for simplicity, we assume that $H=0$. 
Then, for the same $\lambda$, the QSLs are obtained as 
\begin{align*}
     &T_{\ast}(\ket{\Phi^{\pm}}) 
         = \frac{2 \lambda}{\sqrt{5}\gamma}-\frac{2}{5\gamma}\ln(1+ \lambda), \\
     &T_{\ast}(\ket{\Psi^{+}}) 
         = \frac{\lambda}{2\gamma}-\frac{1}{4\gamma}\ln(1+2\lambda). 
\end{align*}
Also we find $T_{\ast}(\ket{\Psi^{-}})\to\infty$ because of $\mathcal{A}=0$, which is equivalent 
to that $\ket{\Psi^{-}}$ is identical to an eigenstate of $M$. 
Thus, $\ket{\Psi^{-}}$ is the most robust Bell state in our definition. 
Moreover, $T_{\ast}(\ket{\Phi^{\pm}}) > T_{\ast}(\ket{\Psi^{+}})$ always holds, and hence 
$\ket{\Psi^{+}}$ is the most fragile state. 
Note that, for the case of the non-collective (local) decoherence modeled by 
$M_{1}=\sqrt{\gamma}\sigma_{-}\otimes I$ and $M_{2}=\sqrt{\gamma}I\otimes \sigma_{-}$, 
we have $\mathcal{A}=\sqrt{5}\gamma$ and $\mathcal{E}=\gamma$ for all Bell states. 
That is, in this case there is no difference of states in robustness.


\subsection{Atomic ensemble}

Next let us consider an ensemble composed of $N$ identical atoms. 
As typical states, we consider the product state of superposition 
$\ket{+}^{\otimes N}=(\ket{0}/\sqrt{2}+\ket{1}/\sqrt{2})^{\otimes N}$ and the GHZ state 
$\ket{{\rm GHZ}}=(\ket{0}^{\otimes N}+\ket{1}^{\otimes N})/\sqrt{2}$; the latter is a 
powerful resource in quantum metrology such as the frequency standard. 
In fact, $\ket{{\rm GHZ}}$ enables us to estimate the frequency with error (standard deviation) 
of the order $1/N$, while $1/\sqrt{N}$ is the best order in the case of $\ket{+}^{\otimes N}$ 
\cite{Bollinger}.
However, in a realistic situation, the system is always subjected to noise, typically the dephasing 
noise $M=\sqrt{\gamma}\sum^{N}_{j=1}\sigma_{z}^{(j)}$ where $\sigma_{z}^{(j)}$ acts on the 
$j$th atom, which vanishes the quantum advantage unless a specific control is applied 
\cite{Huelga}.

To understand this undesired effect brought by the dephasing noise in the language 
of QSL, let us examine $T_*$ of those two states. 
For simplicity, we assume that the magnitude of the system Hamiltonian $H$ is much 
smaller than $\gamma$. 
Then we have 
\begin{eqnarray*}
     (\mathcal{A}, \mathcal{E})(\ket{+}^{\otimes N})
         & \approx &\left(\gamma\sqrt{6N^{2}-2N}, \ \gamma N\right), \\
     (\mathcal{A}, \mathcal{E})(\ket{{\rm GHZ}}) 
         &\approx& \left(2\gamma N^{2},\ \gamma N^{2}\right).
\end{eqnarray*}
These expressions lead to $T_{\ast}(\ket{+}^{\otimes N})\sim O(1/N)$ and 
$T_{\ast}(\ket{{\rm GHZ}})\sim O(1/N^{2})$, for fixed $\lambda$ and $\gamma$. 
Therefore, although $\ket{{\rm GHZ}}$ can ideally improve the estimation accuracy, it is 
more fragile than $\ket{+}^{\otimes N}$.


\section{Hamiltonian engineering for robust state preparation}

In the examples of Sec.~\ref{sec:examples}, we identified a robust state 
$\rho_0=\ket{\psi_0}\bra{\psi_0}$, under given decoherence and system Hamiltonian. 
In this section, in contrast, we discuss designing an optimal Hamiltonian 
that maximizes $T_*$, given a fixed initial state $\rho_0$ and decoherence. 
That is, we aim to find $H$ that protects $\rho_0$ against a given decoherence $M$ by 
maximizing the lower bound of the escaping time of state from a region centered at $\rho_0$.

Now, the problem of maximizing $T_*$ with respect to $H$ is equivalent to that of 
minimizing $\mathcal{A}$, because $T_{\ast}$ is a monotonically decreasing function 
with respect to $\mathcal{A}$ (see Appendix B). 
In particular, we consider the following cost function: 
\begin{align*}
    F(H) = {\rm Tr}(H^2\rho_{0})-{\rm Tr}(H\rho_{0}H\rho_{0})
                +{\rm Tr}(i[\rho_{0},  \mathcal{D}^{\dagger}[M]\rho_0]H),
\end{align*}
which is identical to $4\mathcal{A}^{2}-2{\rm Tr}\left[ (\mathcal{D}^{\dagger}[M]\rho_0)^2\right]$. 
Note that, as clearly seen from the above expression, $F(H)$ is a convex quadratic 
function with respect to $H$. 
Hence, the optimal $H_{\rm opt}$ can be effectively determined. 
Note that this easy-to-handle problem can be formulated thanks to the explicit expression 
of the QSL; recall that this was the motivation to derive $T_*$ and compare it to $T_{\rm DC}$.

In order to have $H_{\rm opt} = {\rm arg min}_H F(H)$, let us take the derivative of $F(H)$ 
with respect to $H$:
\begin{align*}
     \frac{\partial F}{\partial H}
         =(H\rho_0+\rho_0 H)^{\top}-2(\rho_0H\rho_0)^{\top}
             + i\left([\rho_0, \mathcal{D}^{\dagger}[M]\rho_0] \right)^{\top},
\end{align*}
where we have used the following matrix formulae \cite{matrix}: 
\begin{align*}
     & \frac{\partial}{\partial X} {\rm Tr}(XA) = A^{\top}, ~~
         \frac{\partial}{\partial X} {\rm Tr}(X^{2}A) =(XA+AX)^{\top}, 
\notag \\
     &  \frac{\partial}{\partial X} {\rm Tr}(AXAX) =2 (AXA)^{\top}.
\end{align*}
Therefore, $H_{\rm opt}$ satisfies 
\begin{align}
\label{H condition}
      H_{\rm opt} \rho_0 + \rho_0 H_{\rm opt} 
          - 2\rho_0 H_{\rm opt} \rho_0 + i[\rho_0, \mathcal{D}^{\dagger}[M]\rho_0]=0.
\end{align}
This is a simple linear equation with respect to $H$, for a given $\rho_0$ and $M$. 
Thus, $H_{\rm opt}$ can be effectively computed by solving Eq.~\eqref{H condition}.


\subsection{Qubit example}

Let us again consider the qubit subjected to the decoherence $M=\sqrt{\gamma}\sigma_{-}$. 
Also we choose the initial state to be protected as $\ket{\psi_{0}}=[1/2, \sqrt{3}/2]$. 
We represent $H_{\rm opt}$ as $H_{\rm opt}=u_{1}\sigma_{x}+u_{2}\sigma_{y}+u_{3}\sigma_{z}$, 
where $(u_1, u_2, u_3)$ are real parameters to be determined. 
Then by solving the linear equation \eqref{H condition} we have 
\begin{align*}
    u_{2}=-\frac{\sqrt{3}}{16}\gamma, \  u_1+\sqrt{3}u_{3}=0.
\end{align*}
The term $u_1\sigma_x+u_3\sigma_z$ always commutes with 
$\rho_0=\ket{\psi_0}\bra{\psi_0}$ when $u_1+\sqrt{3}u_{3}=0$. 
Thus, only the $u_2\sigma_y$ term has an effect on the dynamics of $\Theta_t$ given in 
Eq.~\eqref{theorem proof eq 1}. 
Figure~\ref{H engineering figure}(a) shows the time evolution of $\cos\Theta_{t}$, in the 
following three cases: $H=H_{\rm opt}$, $H=0$ (i.e., the system is purely decohered), 
and $H=\sigma_z$. 
The decoherence strength is chosen as $\gamma=1$. 
Clearly, $H_{\rm opt}$ makes longer the time for the state escaping from 
${\mathcal R}_\lambda(\rho_0)$ for any $\lambda=\sqrt{1-\cos\Theta_T}$, than the other 
two cases. 
In particular, when $H_{\rm opt}$ is applied, the state remains in the region 
${\mathcal R}_\lambda(\rho_0)$ with radius $\lambda=\sqrt{1-0.9}$, for all time.


\subsection{Qutrit example}

\begin{figure}[tb]
\includegraphics[width=8.7cm]{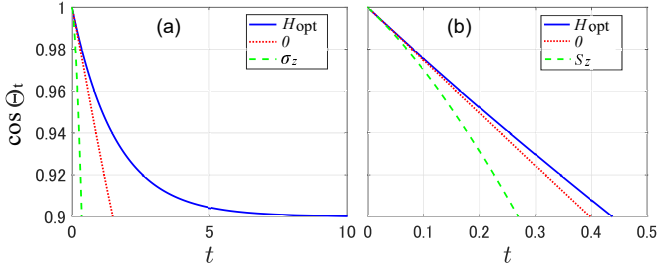}
\caption{Time evolution of $\cos \Theta_{t}$ in the case of (a) qubit driven by 
$H=H_{\rm opt}$ (blue solid line), $0$ (red dotted line), and $\sigma_{z}$ (green dashed line). 
Also the case of (b) qutrit driven by $H=H_{\rm opt}$ (blue solid line), $0$ (red dotted line), 
and $S_{z}$ (green dashed line). }
\label{H engineering figure}
\end{figure}

Lastly, we study a qutrit system, composed of the three orthogonal states $\ket{E}=[1, 0, 0]^{\top}$, 
$\ket{S}=[0, 1, 0]^{\top}$, and $\ket{G}=[0, 0, 1]^{\top}$. 
We assume that the system is subject to the decoherence 
$M=\sqrt{\gamma}(\ket{S}\bra{E}+\ket{G}\bra{S})$, which induces the ladder-type decay 
$\ket{E}\to\ket{S}\to\ket{G}$. 
The target initial state is chosen as $\ket{\psi_{0}}=[1/2, 1/\sqrt{2}, 1/2]^{\top}$. 
The control Hamiltonian to be determined can be parametrized as 
$H_{\rm opt}=\sum^{8}_{i=1}u_{i}\Lambda_{i}$, where $\{ \Lambda_{i}\}^{8}_{i=1}$ are the 
Gell-Mann matrices given in Appendix C. 
In this setting, using Eq.~\eqref{H condition}, we obtain the condition 
\begin{align*}
\begin{cases}
&2\sqrt{2}u_1+5u_3+2u_4-2\sqrt{2}u_6+\sqrt{3}u_8=0, \\
&3u_3+2u_4-\sqrt{3}u_8=0, \\
&3\sqrt{2}u_2+2u_5-\sqrt{2}u_7 =-\gamma,\\
&8u_5+8\sqrt{2}u_7=-\gamma.
\end{cases}
\end{align*}
Under this condition, the terms with coefficients $(u_1, u_3, u_4, u_6, u_8)$ in $H_{\rm opt}$ 
always commute with $\rho_0$ and thus they do not affect on the dynamics of $\Theta_t$ 
as well as ${\mathcal A}$. 
Figure~\ref{H engineering figure}(b) shows the time evolution of $\cos\Theta_{t}$, in the 
cases of $H=H_{\rm opt}$ with the parameters 
$(u_2, u_5, u_7) = (-3\gamma/8\sqrt{2}, 0, -\gamma/8\sqrt{2})$ and compares it to the 
cases $H=0$ and $S_{z}=\ket{E}\bra{E}-\ket{G}\bra{G}$. 
We find that certainly $H=H_{\rm opt}$ makes the escaping time longer, though the 
advantage over the other two cases is not so big compared to the previous qubit example.

\section{Conclusion}

In this paper, we posed an idea to use the QSL to characterize robust quantum states 
and, based on it, formulated the engineering problem of a Hamiltonian that makes a target 
state robust against a given decoherence. 
In this engineering problem it is important for the QSL to be explicitly computable; the new 
QSL derived in this paper indeed satisfies this condition, and further, it is tighter than 
another known QSL in a setup where the robustness issue is critical (that is, the case where 
the decoherence is small and the region ${\mathcal R}_\lambda(\rho_0)$ is small). 
In addition, the Hamiltonian engineering problem is proven to be a convex quadratic 
optimization problem, which is efficiently solvable. 
Several examples have been studied, especially showing another view on the fragility of GHZ 
state in the quantum metrology. 
We hope that the results given in this paper will provide a new perspective of QSL as a tool 
in quantum engineering.


\appendix

\section{$T_{\ast}$ as a function of the decoherence strength}

We here prove that $T_{\ast}$ monotonically decreases with respect to the strength 
of the decoherence, $\gamma$, which is defined through $M=\sqrt{\gamma}M'$ 
with fixed $M'$. 
In terms of $\gamma$, we can express $\mathcal{A}$ and $\mathcal{E}$ as 
$\mathcal{A}=\sqrt{a\gamma^{2}+b\gamma+c}$ and $\mathcal{E}=d\gamma$, where 
$(a, c, d)$ are non-negative constants and $b$ is a constant. 
Then $T_{\ast}$ can be written as 
\begin{align*}
     T_{\ast} &=\frac{2\lambda}{\sqrt{a\gamma^{2}+b\gamma+c}} \\
        &\ \ +\frac{d\gamma}{a\gamma^{2}+b\gamma+c}
              \ln\left(\frac{d\gamma}{d\gamma+\lambda\sqrt{a\gamma^{2}+b\gamma+c}}\right), 
\end{align*}
and $\partial T_{\ast}/\partial \gamma$ is calculated as
\begin{align*}
     \frac{\partial T_{\ast}}{\partial \gamma} 
&=-\frac{(2a\gamma+b)\lambda}{\mathcal{A}^{3}}
    +\frac{2d(a\gamma^{2}-c)}{\mathcal{A}^{4}}
      \ln\left(1+\frac{\mathcal{A}\lambda}{\mathcal{E}} \right) \\
&\ \ \ +\frac{d(b\gamma+2c)\lambda}{\mathcal{A}^{3}(\mathcal{E}+\mathcal{A}\lambda)}.
\end{align*}
Our goal is to show $\partial T_{\ast}/\partial \gamma \leq 0$. 
The proof is divided into three cases: $a\gamma^{2}>c$, $a\gamma^{2}=c$, and 
$a\gamma^{2}<c$. 
First, for the case $a\gamma^{2}>c$ we have 
\begin{align*}
\frac{\partial T_{\ast}}{\partial \gamma} 
&\leq -\frac{(2a\gamma+b)\lambda}{\mathcal{A}^{3}}+\frac{2d(a\gamma^{2}-c)}{\mathcal{A}^{4}} 
\frac{\mathcal{A}\lambda(\mathcal{A}\lambda+2\mathcal{E})}{2\mathcal{E}( \mathcal{A}\lambda+\mathcal{E})} \\
&\ \ +\frac{d(b\gamma+2c)\lambda}{\mathcal{A}^{3}(\mathcal{E}+\mathcal{A}\lambda)} 
=\frac{-\lambda^{2}}{\gamma(\mathcal{E}+\mathcal{A}\lambda )} \leq 0, 
\end{align*}
where the inequality $\ln(1+x) \leq x(x+2)/2(x+1)$ for $x\geq0$ is used. 
Next, for the case $a\gamma^{2}=c$, 
\begin{align*}
\frac{\partial T_{\ast}}{\partial \gamma}  
=-\frac{\lambda^{2}}{\gamma(\mathcal{E}+\mathcal{A} \lambda)} \leq 0.
\end{align*}
Lastly, for the case $a\gamma^{2}<c$, we have 
\begin{align*}
\frac{\partial T_{\ast}}{\partial \gamma} 
&= -\frac{(2a\gamma+b) \lambda}{\mathcal{A}^{3}}+\frac{2d(a\gamma^{2}-c)}{\mathcal{A}^{5}\lambda} \mathcal{A}\lambda\ln\left(\frac{\mathcal{E} +\mathcal{A}\lambda}{\mathcal{E}} \right) \\
&\ \ +\frac{d(b\gamma+2c) \lambda}{\mathcal{A}^{3}(\mathcal{E}+\mathcal{A} \lambda)} \\
&\leq -\frac{(2a\gamma+b) \lambda}{\mathcal{A}^{3}}+\frac{2d(a\gamma^{2}-c)}{\mathcal{A}^{5}\lambda} \frac{2\mathcal{A}^{2}\lambda^{2}}{2\mathcal{E}+\mathcal{A}\lambda} \\
&\ \ +\frac{d(b\gamma+2c) \lambda}{\mathcal{A}^{3}(\mathcal{E}+\mathcal{A} \lambda)} \\
&=\frac{-\lambda^{2}}{\mathcal{A}^{2}(\mathcal{E}+\mathcal{A}\lambda)}\left(2a\gamma+b+\frac{2d(c-a\gamma^{2})}{2\mathcal{E}+\mathcal{A}\lambda}  \right),
\end{align*}
where the inequality $2(\alpha-\beta)^{2}/(\alpha+\beta) \leq (\alpha-\beta)\ln(\alpha/\beta)$ 
for $\alpha, \beta \geq 0$ is used. 
Now, $2a\gamma+b>0$ readily leads to $\partial T_{\ast}/\partial \gamma \leq 0$. 
If $2a\gamma+b\leq 0$, the above inequality can be further computed as 
\begin{align*}
\frac{\partial T_{\ast}}{\partial \gamma} &\leq \frac{-\lambda^{2} (2a\gamma^{2}+b\gamma)(2\mathcal{E}+\mathcal{A}\lambda)}{\gamma\mathcal{A}^{2}(\mathcal{E}+\mathcal{A}\lambda)(2\mathcal{E}+\mathcal{A}\lambda) } \\
&\ \ +\frac{2\lambda^{2}(a\gamma^{2}-c)}{\gamma\mathcal{A}^{2}(\mathcal{E}+\mathcal{A}\lambda)(2\mathcal{E}+\mathcal{A}\lambda) } \\
&=\frac{-\lambda^{2}\{2\mathcal{E}\mathcal{A}^{2}-(4\mathcal{E}+\mathcal{A}\lambda)(2a\gamma^{2}+b\gamma)\}}{\gamma\mathcal{A}^{2}(\mathcal{E}+\mathcal{A}\lambda)(2\mathcal{E}+\mathcal{A}\lambda) } \leq 0.
\end{align*}


\section{$T_{\ast}$ as a function of $\mathcal{A}$}

We here prove that $T_{\ast}$ is a monotonically decreasing function with respect to 
$\mathcal{A}$, as follows. First, 
\begin{align*}
    \frac{\partial T_{\ast}}{\partial \mathcal{A}} 
       &=-\frac{2}{\mathcal{A}^{2}} 
              \left\{ \lambda+\frac{\mathcal{E}}{\mathcal{A}}
                  \ln \left(\frac{\mathcal{E}}{\mathcal{E}+\mathcal{A} \lambda}\right) \right\} \\
       &\ \ \ -\frac{2\mathcal{E}}{\mathcal{A}^{2}}
               \left\{ \frac{1}{\mathcal{A}}
                   \ln \left( \frac{\mathcal{E}}{\mathcal{E}+\mathcal{A} \lambda}\right) 
               + \frac{ \lambda}{\mathcal{E}+\mathcal{A} \lambda} \right\} \\
       & = -\frac{2}{\mathcal{A}^{2}}
                \left\{\lambda + \frac{k \lambda}{k+\lambda}
                    -2k\ln \left(1+ \frac{\lambda}{k} \right)\right\},
\end{align*}
where $k=\mathcal{E}/\mathcal{A}$. 
Then from the inequality $\ln(1+x) \leq x(2+x)/2(1+x)$ for $x\geq0$, we have 
\begin{align*}
      \frac{\partial T_{\ast}}{\partial \mathcal{A}} 
         \leq -\frac{2}{\mathcal{A}^{2}}
             \left\{ \lambda + \frac{k\lambda}{k+\lambda}
                 - 2k \frac{ (\lambda/k) (2+\lambda/k)}{2(1+\lambda/k)} \right\}=0.
\end{align*}


\section{Gell-Mann matrices}

The Gell-Mann matrices $\{ \Lambda_{i}\}^{8}_{i=1}$ are defined as
\begin{align*}
\Lambda_{1}&= \left[
    \begin{array}{rrr}
      0 & 1& 0  \\
      1 & 0  &0 \\
     0 & 0& 0
    \end{array}
  \right], \ 
\Lambda_{2}= \left[
    \begin{array}{rrr}
      0 & -i& 0  \\
      i & 0  &0 \\
     0 & 0& 0  
    \end{array}
  \right], \ 
\Lambda_{3}= \left[
    \begin{array}{rrr}
      1 & 0& 0  \\
      0 & -1  &0 \\
     0 & 0& 0
    \end{array}
  \right], \\
\Lambda_{4}&= \left[
    \begin{array}{rrr}
      0 & 0& 1  \\
      0 & 0  &0 \\
     1 & 0& 0
    \end{array}
  \right], \ 
\Lambda_{5}= \left[
    \begin{array}{rrr}
      0 & 0& -i  \\
      0 & 0  &0 \\
      i & 0& 0  
    \end{array}
  \right], \ 
\Lambda_{6}= \left[
    \begin{array}{rrr}
      0 & 0& 0  \\
      0 & 0  &1 \\
     0 & 1& 0
    \end{array}
  \right], \\
\Lambda_{7}&= \left[
    \begin{array}{rrr}
      0 & 0& 0  \\
      0 & 0  &-i \\
      0 & i& 0  
    \end{array}
  \right], \ 
\Lambda_{8}= \frac{1}{\sqrt{3}}\left[
    \begin{array}{rrr}
      1 & 0& 0  \\
      0 & 1  &0 \\
     0 & 0& -2
    \end{array}
  \right].
\end{align*}
Note that these matrices form an orthonormal basis set in {\it SU}(3).



\begin{thebibliography}{}



\bibitem{computation1}
S. Lloyd, 
Ultimate physical limits to computation, 
Nature {\bf 406}, 1047 (2000).

\bibitem{computation2}
M. A. Nielsen and I. L. Chuang, 
{\it Quantum Computation and Quantum Information} 
(Cambridge University Press, Cambridge, 2010). 

\bibitem{metrology1}
P. J. Jones and P. Kok, 
Geometric derivation of the quantum speed limit, 
Phys. Rev. A {\bf 82}, 022107 (2010).

\bibitem{metrology2}
F. Frowis, 
Kind of entanglement that speed up quantum evolution, 
Phys. Rev. A {\bf 85}, 052127 (2012).

\bibitem{opt1}
T. Caneva, M. Murphy, T. Calarco, R. Fazio, S. Montangero, V. Giovannetti, and G. E. Santoro, 
Optimal Control at the Quantum Speed Limit, 
Phys. Rev. Lett. {\bf 103}, 240501 (2009).


\bibitem{opt2}
P. M. Poggi, F. C. Lombardo, and D. A. Wisniacki, 
Quantum speed limit and optimal evolution time in a two-level system, 
EPL {\bf 104} (2013).

\bibitem{opt3}
O. Andersson and H. Heydari, 
Quantum speed limits and optimal Hamiltonians for driven systems in mixed states, 
J. Phys. A: Math. Theor. {\bf 47}, 215301 (2014).




\bibitem{MT}
L. Mandelstam and I. Tamm, 
The uncertainty relation between energy and time in nonrelativistic quantum mechanics, 
J. Phys. (USSR) {\bf 9}, 249-254 (1945). 


\bibitem{Uhlmann}
A. Uhlmann, 
An energy dispersion estimate, 
Phys. Lett. A {\bf 161}, 329 (1992). 


\bibitem{Pfeifer1}
P. Pfeifer, 
How Fast Can a Quantum State Change with Time?, 
Phys. Rev. Lett. {\bf 70}, 22, 3365 (1993).


\bibitem{Pfeifer2}
P. Pfeifer and J. Frohlich, 
Generalized time-energy uncertainty relations and bounds on lifetimes of resonance, 
Rev. Mod. Phys. {\bf 67}, 759 (1995).

\bibitem{Deffner2}
S. Deffner and E. Lutz, 
Energy-time uncertainty relation for driven quantum systems, 
J. Phys. A {\bf 46}, 335302 (2013). 



\bibitem{ML}
N. Margolus and L. B. Levitin, 
The maximum speed of dynamical evolution, 
Physica D {\bf 120}, 188 (1998).


\bibitem{Taddei}
M. M. Taddei, B. M. Escher, L. Davidovich, and R. L. de Matos Filho, 
Quantum Speed Limit for Physical Processes, 
Phys. Rev. Lett. {\bf 110}, 050402 (2013).


\bibitem{Campo}
A. del Campo, I. L. Egusquiza, M. B. Plenio, and S. F. Huelga, 
Quantum Speed Limits in Open System Dynamics, 
Phys. Rev. Lett. {\bf 110}, 050403 (2013).


\bibitem{Deffner}
S. Deffner and E. Lutz, 
Quantum Speed Limit for Non-Markovian Dynamics, 
Phys. Rev. Lett. {\bf 111}, 010402 (2013).

\bibitem{Sun}
Z. Sun, J. Liu, J. Ma, and X. Wang, 
Quantum speed limits in open systems: Non-Markovian dynamics without rotating-wave 
approximation, 
Sci. Rep. {\bf 5}, 8444 (2015). 


\bibitem{Meng}
X. Meng, C. Wu, and H. Guo, 
Minimal evolution time and quantum speed limit of non-Markovian open systems, 
Sci. Rep. {\bf 7}, 15046 (2015).


\bibitem{Zhang}
Y.-J. Zhang, W. Han, Y.-J. Xia, J.-P Cao, and H. Fan, 
Quantum speed limit for arbitrary initial states, 
Sci. Rep. {\bf 4}, 27349 (2016).


\bibitem{Campaioli}
F. Campaioli, F. A. Pollock, and K. Modi, 
Tight, robust, and feasible quantum speed limits for open dynamics, 
Quantum {\bf 3},168 (2019).


\bibitem{speed1}
Z.-Y. Xu, S. Luo, W. L. Yang, C. Liu, and S. Zhu, 
Quantum speedup in a memory environment, 
Phys. Rev. A {\bf 89}, 012307 (2014). 

\bibitem{speed2}
C. Liu, Z.-Y. Xu, and S. Zhu, 
Quantum-speed-limit for multiqubit open systems, 
Phys. Rev. A  {\bf 91}, 022102 (2015).

\bibitem{speed3} 
S.-X. Wu, Y. Zhang, C.-S. Yu, and H.-S. Song, 
The initial-state dependence of the quantum speed limit, 
J. Phys. A {\bf 48}, 045301 (2015). 


\bibitem{speed4}
Y.-J. Zhang, W. Han, Y.-J. Xia, J.-P. Cao, and H. Fan, 
Classical-driving-assisted quantum speed-up, 
Phys. Rev. A {\bf 91}, 032112 (2015). 


\bibitem{speed5}
Y.-J. Song, Q.-S. Tan, and L.-M. Kuang, 
Control quantum evolution speed of a single dephasing qubit for arbitrary initial states 
via periodic dynamical decoupling pulses, 
Sci. Rep. {\bf 7}, 43654 (2017).


\bibitem{Bollinger}
J. J. Bollinger, W. M. Itano, D. J. Wineland, and D. J. Heinzen, 
Optimal frequency measurements with maximally correlated states, 
Phys. Rev. A {\bf 54}, 4649 (1996). 

\bibitem{Huelga}
S. F. Huelga, C. Macchiavello, T. Pellizzari, A. K. Ekert, M. B. Plenio, and J. I. Cirac, 
Improvement of Frequency Standards with Quantum Entanglement, 
Phys. Rev. Lett. {\bf 79}, 3865 (1997).  

\bibitem{matrix}
K. B. Petersen and M. S. Pedersen, 
The Matrix Cookbook, 
url=http://matrixcookbook.com (2012).



\end{thebibliography}
\end{document}